# Plane and parabolic solar panels

**Jorge Henrique O. Sales**, Instituto de Física Teórica – UNESP, Rua Dr. Bento Teobaldo Ferraz 271, Bloco II, Barra Funda, São Paulo – SP - CEP: 01140-070,
jorgehenrique@unifei.edu.br

**Alfredo T. Suzuki**, Instituto de Física Teórica – UNESP, Rua Dr. Bento Teobaldo Ferraz 271, Bloco II, Barra Funda, São Paulo – SP - CEP: 01140-070,
Suzuki@ift.unesp.br

*Abstract:* We present a plane and parabolic collector that absorbs radiant energy and transforms it in heat. Therefore we have a panel to heat water. We study how to increment this capture of solar beams onto the panel in order to increase its efficiency in heating water.

**1. Introduction**

Man lives today in a time of exhaustion of the planet's natural resources and at the same time in which returns to nature the non recyclable trash of diverse products. Among these, the prominent materials, more aggressive and abundant are such as synthetic, metals (including heavy ones), plastics, and chemicals.

The conscientious awakening that something very serious has to be done is unquestionable and irreversible. Nowadays we think a lot in using clean energy sources like solar, wind, from hydrogen, and electric energy forms, among others. The preoccupation in developing alternative energy forms has its basis not only as an environmental issue, but also as a search for economic alternatives is also fundamental.

In this work we present two low cost models for water heater that meets the environmental and economic needs. We have a plane and a parabolic collector that absorbs solar radiant energy and transforms it in heat. These collectors are important to reduce the electric energy costs since one of the most expensive items for electric energy consumption is in heating water. Specifically for the plane collector we study how to increment the capture of solar beams in the panel to increase its efficiency.

**2. Plane solar panel**

The plane panels (collectors) are boxes with a semi transparent covering (transparent for solar radiation and opaque for infrared radiation) to take advantage of greenhouse effect. Within the box we put a radiator covered by a selective surface that absorbs solar light and heats itself. Collector's bottom and side walls are covered by thermal insulators to avoid heat loss by conduction between the radiator and walls or roof wherein the system is fixed.

Our collectors are made with PVC lining plaques. The absence of the usual glass coverings used in the traditional collectors avoids our collector to heat the water as much as the traditional ones. This temperature reduction however allows the use of PVC ducts for cool water and better still, makes a safer water for children, since too hot a water may cause accidental burns in users, be them an adult or a child. The use of PVC ducts also reduces the cost of these collectors.

The materials that were used for our collectors are:

- Ducts and connections in PVC
- 1" Glove for rigid electro duct for nipple
- 1" white nipple, with external screw, 25 mm internal diameter
- 25 mm flexible electro duct in yellow PVC
- Alveolar backing
- 1250 mm x 620 mm x 10 mm (modular) PVC
- Mercury thermometer graded from -10 $^0$C to +110 $^0$C
- Professional 24 hour Araldite glue.

**2.1. Results**

Cold water that comes from the water reservoir installed in residences enters the plane collector through the PVC ducts. Solar beams hit the plaques full of water and radiating energy is transformed into heat that elevates the water temperature. This heated water then comes out of the collector ready to be used in the manifold places within the home.

Our plane collector has the capacity to heat the water nearly to body temperature (98 $^0$F ~ 37 $^0$C) in a sunny winter day whose maximum temperature was 77 $^0$F ~ 25 $^0$C. (See graph below) (Silva, 2008).

In Figure 1 we have the measured results for water temperature (red curve) and room temperature (blue curve) from 8 o'clock a.m. through 6 o'clock p.m. on June 08, 2008 (winter season) in the city of Itajubá, MG – Brazil. Temperature readings were made with a mercury thermometer in the water that emerged from the three plane plaques of the solar panel coupled to a 200 liters water tank, whose temperature started at 62 $^0$F ~ 17 $^0$C around 7:30 a.m.

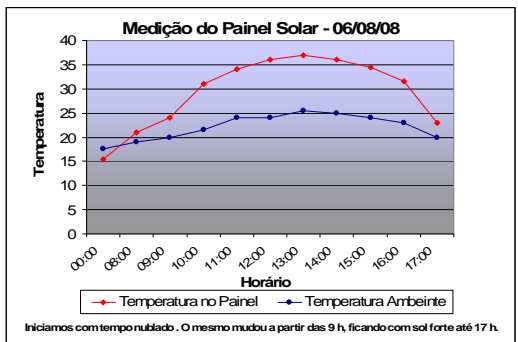

Figure 1. Graph temperature x day hour

Looking at the graph we can see that the most efficient period for heating was between 10 a.m. through 4 p.m. where the curve exhibits a roughly rounded plateau.

With some modifications that will be studied, we expect to enlarge this band to between 7 a.m. through 5 p.m. for a sunny day, thus increasing the efficiency in the capture of solar radiant energy to heat water.

One possibility is to introduce parabolic collectors at the plane panel laterals according to figure 3. They will function as transmitters of radiation to the plane panels.

### 3. Parabolic solar panel

Heat transfer between two bodies happens in three manners: radiation, conduction and convection. Of course, heat transfer may involve more than one of these forms at the same time. (IVAN, R. 1990).

Conduction is the functioning principle for conventional solar heaters found in the market. Our model here presented differs from the competitors by the material employed and by the easiness in building it. In the case for this model, besides the three previous forms, we used a reflecting surface in a parabolic form, so as to concentrate the solar beams onto a small region called focus where for this reason becomes subject to higher solar energy intensities. With this collector it is possible to achieve higher water temperatures than those obtained with the plane collectors.

This kind of collector does not directly "store" solar energy; it depends on a reflecting surface such as an aluminum plaque shaped in a parabolic form as shown by Faria P. et al, 2007.

To serve as this reflecting surface we made use of the internal part of aluminum beer and other soft drinks can. To build a parabolic collector or a cylinder-parabolic, it is necessary in the first place to make a mold of this curve named parabola in the cans. It is in the focus of the parabola that the solar beams are concentrated and where we position the water duct to be heated.

#### 3.1. Model

The aluminum can collectors (as shown in figure 2) have approximate dimensions of 1,1 x 0,9 m in the form of a box closed at the basis with narrow plates of compensated wood around it painted in black and at the top covered by glass. This glass covering produces the greenhouse effect resulting in a better efficiency for the system. Side by side aligned cans along its axial axis form the set of parabolic collectors.

In Figure 2 we have:
1. 50 mm x 30 mm x 1.5 mm steal cupboard corner;
2. Aluminum can (cut in the middle);
3. $90^0$ PVC elbow Ø 1/2";
4. 6 mm blackened plywood panel;
5. PVC duct Ø 1/2";
6. 4 mm transparent acrylic panel;
7. PVC duct Ø 1/2";
8. PVC duct Ø 1/2".

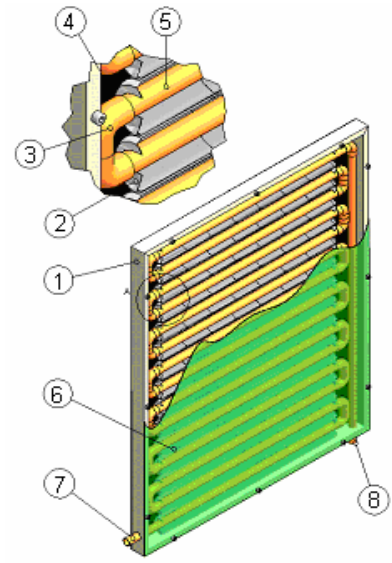

Figure 2. Parabolic energy accumulator

#### 3.2. Results

Cold water coming from the water tank installed in residences enters the aluminum can collector through the PVC ducts, see figure 2. Solar beams are reflected by these cans molded in the form of a parabola and focused onto the PVC ducts full of water which then becomes heated.

Our cylinder-parabolic energy collector heats the water inside the ducts up to 176 $^0$F ~ 80 $^0$C in a hot sunny day, with day temperatures around 86 $^0$F ~ 30 $^0$C. This panel is capable of heating 12 liters of water (approximate water consumption per shower per person) that enters the duct at 77 $^0$F ~ 25 $^0$C.

The use of this kind of solar collector by the low income families allows an average 35% reduction in electric energy, which for the Brazilian standards for costs in electricity, corresponds to a reduction of 56% to 71% in the electric energy spending by the families (Faria, 2007).

**4. Plane-parabolic solar panel**

We are studying now the viability of increasing the efficiency of light capture for a period before 7 a.m. and after 5 p.m., that is, average Brazilian sunrise and sunset thresholds. We expect that this will help increasing water temperature by 10% from the present stage as shown in Figure 1.

One possibility is to couple parabolic collectors alongside the laterals of the plane panels, as in Figure 3. This will allow additional focusing of solar radiation onto the plane panels.

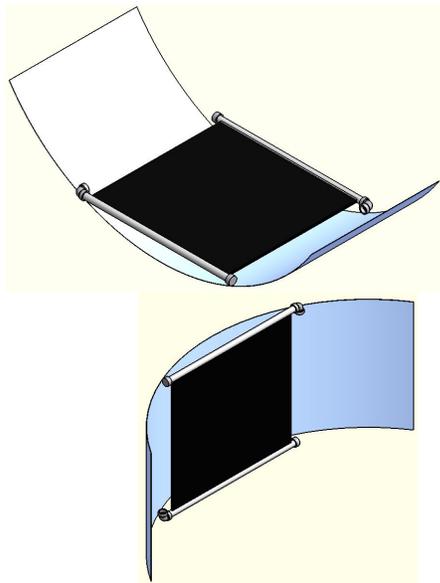

Figure 3. Plane-parabolic solar panel.

**5. Conclusion**

Plane plaque energy accumulator is in its final tests to be implemented in needy residences for low income population in the city of Itajubá-MG – Brazil. Parabolic panel is patented at INPI and solar plane-parabolic panel is in test phase using a prototype built in Pro-E.

First results on water temperature and period of keeping the water heated show up to the moment an efficiency of 60% when compared to the electric shower.

**6. Acknowledgments**

To FAPESP, FAPEMIG, Associação Ecológica Amigos do Rio Sapucaí de Itajubá-AEARSI. FUNDUNESP for partial financial support and to Instituto de Física Teórica de São Paulo-UNESP for hospitality.

**7. References**

Faria, P. H; Verraci, M. H. ; Sales, J. H. O. Boletim Técnico da FATEC-SP - **23** – pág. 104 (2007).

Ivan, R.; Toledo, N. Os Fundamentos da Física Volume 2. Editora Moderna 1990.

Silva, F.C; Moreira, A.P; Andrade, G.; Verraci, M. H.; Sales, J. H. O.; Suzuki, A. T. Boletim Técnico da FATEC-SP - **25** – pág. 77 (2008).